\RequirePackage{fix-cm}

\documentclass[article]{IEEEtran}

\usepackage{algorithm}
\usepackage{algorithmic}
\usepackage{graphicx}
\usepackage{amsmath}
\usepackage{array}
\usepackage{subfigure}
\newcolumntype{x}[1]{%
>{\raggedleft}b{#1}}%

\newcommand{\codelen}{n} %Was N
\newcommand{\codeloglen}{m} %Was n
\newcommand{\bfu}{\mathbf{u}} %Was U
\newcommand{\bfc}{\mathbf{c}} %Was X
\newcommand{\bfy}{\mathbf{y}} %Was Y
\newcommand{\regc}{c} %Was x
\newcommand{\nodeProcessorComplexity}{C_\mathrm{np}}
\newcommand{\nodeProcessorTime}{t_\mathrm{np}}
\newcommand{\registerComplexity}{C_\mathrm{r}}
\newcommand{\muxComplexity}{C_\mathrm{mux}}
\newcommand{\myvec}[1]{(#1)}
\newcommand{\Prob}{\mathrm{Pr}}
\newcommand{\comment}[1]{}

\newcommand{\tn}{\tabularnewline}
 
\usepackage{ifpdf}
\ifpdf
\usepackage[final,pdfborder={0 0 0}]{hyperref}
\fi
\usepackage{color}

\def\Hat{\mathaccent"705E }
\newcommand{\Arikan}{Ar\i kan}
\def\Node{$\cal{N}$}

\begin{document}

\title{Hardware Implementation of Successive Cancellation Decoders for Polar Codes}

%\titlerunning{Hardware Implementation of Successive Cancellation Polar Code Decoders}

\author{\IEEEauthorblockN{Camille Leroux\IEEEauthorrefmark{1}, Alexandre J. Raymond\IEEEauthorrefmark{2}, Gabi Sarkis\IEEEauthorrefmark{2}, Ido Tal\IEEEauthorrefmark{3}, Alexander Vardy\IEEEauthorrefmark{3} and Warren J. Gross\IEEEauthorrefmark{2}}
\IEEEauthorblockA{\IEEEauthorrefmark{1}Institut Polytechnique de Bordeaux, CNRS IMS, UMR 5218, Bordeaux, France.\\}
\IEEEauthorblockA{\IEEEauthorrefmark{2}McGill University, Montreal, QC, Canada.\\}
\IEEEauthorblockA{\IEEEauthorrefmark{3}University of California San Diego, La Jolla, CA, USA.}
}

%\author{\IEEEauthorblockN{Camille Leroux, Alexandre J. Raymond, Gabi Sarkis, Ido Tal, Alexander Vardy and Warren J. Gross}\\

%Institut Polytechnique de Bordeaux, Bordeaux, France\\
%Email: camille.leroux@ibp.fr, \{gabi.sarkis,alexandre.raymond\}@mail.mcgill.ca, warren.gross@mcgill.ca}
%}

%\institute{C. Leroux \at
%  Institut Polytechnique de Bordeaux\\Bordeaux, France\\
%  \email{camille.leroux@ipb.fr}
%  \and
%  A. J. Raymond \at
%  McGill University\\Montreal, QC. Canada\\
%  \email{alexandre.raymond@mail.mcgill.ca}
%  \and
%  G. Sarkis \at
%  McGill University\\Montreal, QC. Canada\\
%  \email{gabi.sarkis@mail.mcgill.ca}
%  \and
%  I. Tal \at
%  University of California San Diego\\La Jolla, CA. USA\\
%  \email{idotal@ieee.org}
%  \and
%  A. Vardy \at
%  University of California San Diego\\La Jolla, CA. USA\\
%  \email{avardy@ucsd.edu}
%  \and
%  W. J. Gross \at
%  McGill University\\
%  3480 University Street\\
%  Montreal, QC. Canada H3A 2A7\\
%  Tel: +1 514 398-2812\\
%  \email{warren.gross@mcgill.ca}
%}

\date{Received: date / Accepted: date}

\maketitle

\begin{abstract}
The recently-discovered polar codes are seen as a major breakthrough in coding theory; they provably achieve the theoretical capacity of discrete memoryless channels using the low complexity successive cancellation (SC) decoding algorithm. Motivated by recent developments in polar coding theory, we propose a family of efficient hardware implementations for SC polar decoders. We show that such decoders can be implemented with $O(n)$ processing elements, $O(n)$ memory elements, and can provide a constant throughput for a given target clock frequency. Furthermore, we show that SC decoding can be implemented in the logarithm domain, thereby eliminating costly multiplication and division operations and reducing the complexity of each processing element greatly. We also present a detailed architecture for an SC decoder and provide logic synthesis results confirming the linear growth in complexity of the decoder as the code length increases.

\end{abstract}

\section{Introduction}
\label{sec:intro}

Polar codes \cite{channel_polarization} form a family of error correcting codes with an explicit and efficient construction \cite{TalVardy} encoding and decoding algorithms. They achieve channel capacity---asymptotically in the code length $n$---when the underlying channel is memoryless and has a discrete input alphabet \cite{Sasoglu2009}. To date, they are the first codes to provably achieve channel capacity with tractable decoding complexity.
Moreover, in some information theoretic applications, such as achieving the secrecy capacity of the wiretap channel in the general case, polar codes are the only known solution which is both explicit and efficient \cite{PC_wiretap}. They are therefore seen as a major breakthrough in coding and information theory.

From a practical point of view, however, polar codes come close to achieving the channel capacity only for very large code lengths, e.g. $\codelen \geq 2^{20}$. Recent works have therefore started to address the issue of performance at shorter code lengths. For example, it was shown in \cite{perf_PC} that the belief propagation (BP) decoding of polar codes improved their performance compared to successive cancellation (SC) decoding without an increase in block length $n$. This performance gain is however obtained at the expense of an increase in decoding complexity. List decoding \cite{tal2011} also improves performance without an increase in code length; however, decoding complexity grows linearly in list size.

Driven by recent theoretical advances related to polar codes and the extra complexity incurred by the use of BP or list decoding, we aim to find efficient hardware architectures for SC decoding, allowing both high throughput and low area implementations of moderate length polar decoders.
Starting from the general framework proposed by \Arikan{} \cite{channel_polarization} and described in Section~\ref{sec:pc}, we develop multiple decoder architectures in order of decreasing hardware complexity and show that SC decoding can actually be implemented with hardware complexity $O(\codelen)$ using the line decoder in Section~\ref{sec:sc-arch}. Finally, We address the implementation of the decoder and its computational nodes and present logic synthesis results confirming our complexity analysis in Section~\ref{sec:hw}.

\section{Polar Codes}
\label{sec:pc}
\begin{figure}[t]
  \centering
  \centerline{\includegraphics[width=7cm]{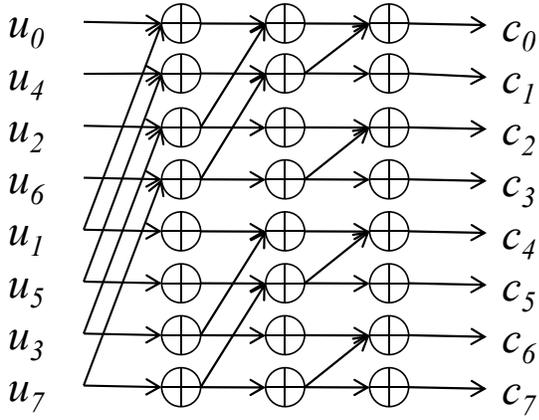}}
  \caption{Encoder architecture for $n=8$.}
\label{fig:encoder}
\end{figure}

A polar code is a linear block error-correcting code designed for a specific discrete input, memoryless channel. From here on, we will assume that the channel has a binary input alphabet and is symmetric as well \cite{Gallager68}.
Let $\codelen = 2^\codeloglen$ be the code length and let $\bfu=\myvec{u_0,u_1,\ldots,u_{\codelen-1}}$ and $\bfc=\myvec{\regc_0,\regc_1,\ldots,\regc_{\codelen-1}}$ denote the input bits and the corresponding codeword\footnote{Note that $n$ input bits are encoded to a length $n$ codeword. However, as we will see later on, not all of the $n$ input bits carry information.}, respectively.
The encoding operation has a Fast-Fourier-Transform-like butterfly structure depicted in Figure~\ref{fig:encoder} for $\codelen = 8$. Note that the ordering of the $u_i$ bits in Figure~\ref{fig:encoder} is according to the bit-reversed order: if we reverse the order of the bits in the binary representation of $i$, we then get the natural ordering.

After $\bfu$ is encoded into $\bfc$, the codeword $\bfc$ is sent over the underlying channel (the channel is used $\codelen$ times). Denote by $\bfy = \myvec{y_0,y_1,\ldots,y_{\codelen-1}}$ the corresponding channel output. We now wish to decode $\bfy$. This is done in terms of a \emph{successive cancellation} decoder. That is, given $\bfy$, we first try to deduce the value of $u_0$, then that of $u_1$, and so forth up until $u_{\codelen-1}$.
We do this as follows. Assume that we are currently at bit $i$ and have already estimated the values of $u_0,u_1,\ldots,u_{i-1}$ to be $\Hat{u}_0,\Hat{u}_1,\ldots,\Hat{u}_{i-1}$. Next, for $b \in \{0,1\}$, denote by $\Prob(\bfy|\Hat{u}_0^{i-1}, u_i=b)$ the probability that $\bfy$ was received, given that $u_0^{i-1} = \Hat{u}_0^{i-1}$, $u_i = b$, and $u_{i+1},u_{i+2},\ldots,u_{n-1}$ are independent random variables with Bernoulli distribution of parameter 0.5. The estimated value $\Hat {u}_i$ is chosen according to:

\begin{equation}
\label{eq:pc-decision}
  \Hat {u}_i = \left \{ \begin{array}{lll}
0 \textrm{ if } \frac{\Prob(\bfy|\Hat {u}_{0}^{i-1}, u_{i}=0)}{\Prob(\bfy|\Hat {u}_{0}^{i-1}, u_{i}=1)}\geq1,\\
1 \textrm{ otherwise}.\\
  \end{array} \right.
\end{equation}

As the code length, $n$, increases, the probability that a bit $u_i$ is correctly decoded, given that all previous bits were correctly decoded, approaches\footnote{This is true for almost all $i$} either 1 or 0.5 as proven in \cite{channel_polarization}. The fraction of bits whose probability of successful decoding approaches 1 tends towards the capacity of the underlying channel as $n$ increases. This information regarding bit reliabilities is used to select a high reliability subset of $\bfu$ to store information bits; while the rest of $\bfu$, called the frozen bit set, is set to a fixed value, assumed to be 0 in this work.
The frozen set is known at the decoder, which sets $\hat{u}_i$ to 0 if it is in the frozen set, and uses Equation~\eqref{eq:pc-decision} otherwise.

\section{Successive Cancellation Decoder Architectures}
\label{sec:sc-arch}

\subsection{Butterfly-based architecture}
\label{sec:sc-arch:butterfly}

\begin{figure}[t]
  \centering
  \centerline{\includegraphics[width=8cm]{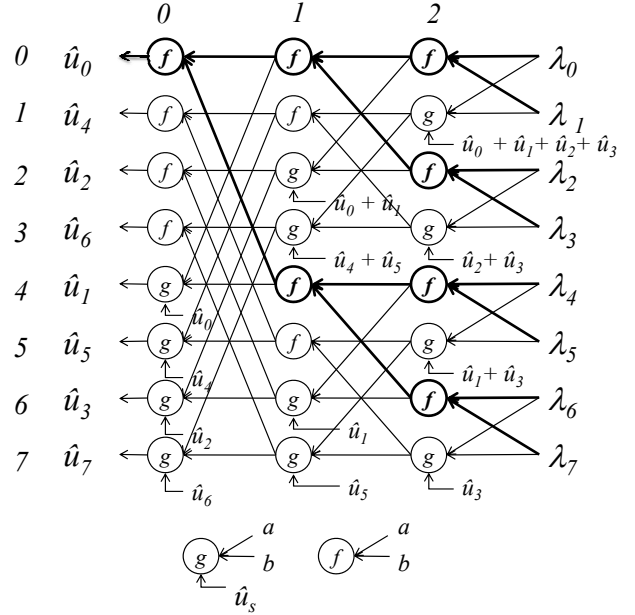}}
  \caption{The butterfly-based SC decoder architecture for $n=8$.}
\label{fig:decoder}
\end{figure}

\Arikan{} showed that SC decoding can be efficiently implemented by the factor graph of the code, which has a structure resembling that of the Fast Fourier Transform. In the remainder of this paper, we will refer to this decoder architecture as the ``butterfly-based SC decoder.'' Figure~\ref{fig:decoder} illustrates the graph of this SC decoder for $\codelen=8$. Channel likelihood ratios (LRs) $\lambda_i$ are assumed to be presented to the right hand side of the graph whereas the estimated bits $\Hat{u}_i$ appear on the opposite end.

The SC decoder is composed of $\codeloglen=\log_2 \codelen$ stages, each containing $\codelen$ nodes. We refer to a specific node as \Node$_{l,j}$ where $l$ designates the stage index $(0\leq l\leq\codeloglen-1)$, and $j$, the node index within stage $l$ $(0\leq j\leq\codelen-1)$. Each node updates its output according to one of the two following update rules:

\begin{equation}
  \begin{split}
    f(a,b) &= \frac{1+ab}{a+b} \textrm{ or}\\
    g_{\Hat{u}_s}(a,b) &= a^{1-2\Hat{u}_s} b.
  \end{split}
\end{equation}

The values $a$ and $b$ are likelihood ratios while $\Hat{u}_s$ is a bit that represents the partial modulo-$2$ sum of previously estimated bits. For example, in node \Node$_{1,3}$, the partial sum is $\Hat{u}_s=\Hat{u}_4 \oplus \Hat{u}_5$. The value of $\Hat{u}_s$ determines if function $g$ should be a multiplication or a division. These update rules are complex to implement in hardware since they involve multiplications and divisions. In Section~\ref{sec:min_sum}, we propose to perform these operations in the logarithm domain and to apply an approximation to the function $f$.

The sequential nature of the algorithm introduces data dependencies to the decoding process. We notice that \Node$_{1,2}$ cannot be updated before bit $\Hat{u}_1$ is computed and a fortiori neither before $\Hat{u}_0$ is known. In order to respect the data dependencies, a scheduling has to be defined. \Arikan{} proposed two schedulings for this decoding framework \cite{channel_polarization}. In the left-to-right scheduling, nodes recursively call their predecessors until an updated node is reached. The recursive nature of this scheduling is especially suitable for software implementation. In the alternative right-to-left scheduling, any node updates its value whenever its inputs are available which enables some nodes to update their values in parallel.
Each bit $\Hat{u}_i$ is successively estimated by activating the spanning tree rooted at \Node$_{0,\pi(i)}$, where $\pi(.)$  denotes the bit-reverse mapping function. As an example, in Figure~\ref{fig:decoder}, the tree associated with $\Hat{u}_0$ is highlighted. If we assume that memory elements are inserted between each stage or equivalently that each node processor can store its updated value, then some results can be reused. For example, in Figure~\ref{fig:decoder}, bit $\Hat{u}_1$ can be decoded by only activating \Node$_{0,4}$ since \Node$_{1,0}$ and \Node$_{1,4}$ have already been updated during the decoding of $\Hat{u}_0$.

Despite this well-defined structure and scheduling of the butterfly-based decoder, in \cite{channel_polarization}, \Arikan{} does not address the problem of resource sharing, memory management or control generation that would be required for hardware implementation. This framework however suggests that it could be implemented with $\codelen \log_2 \codelen$ combinational node processors together with $\codelen$ registers between each stage to store intermediate results. In order to store the channel information, $\codelen$ extra registers are included as well. The total complexity of such a decoder is

\begin{equation}
  C_{butterfly}=(\nodeProcessorComplexity + \registerComplexity)\codelen\log_2 \codelen + \codelen \registerComplexity,
\end{equation}

where $\nodeProcessorComplexity$ and $\registerComplexity$ are the hardware complexity of a node processor and a memory register, respectively.
In order to decode one vector, each stage $l$ has to be activated $2^{m-l}$ times. If we assume that one stage is activated at each clock cycle, then the number of clock cycles required to decode one vector is
\begin{equation}
  NCC=\sum_{l=0}^{m-1}2^{m-l} =2n-2.
  \label{eq:ncc}
\end{equation}

The throughput in bits per second would then be
\begin{equation}
  T=\frac{\codelen}{NCC\times \nodeProcessorTime} \approx \frac{1}{2\nodeProcessorTime},
  \label{eq:throughput}
\end{equation}
where $\nodeProcessorTime$ is the propagation time in seconds through a node processor which also represents the clock period.
It follows that every node processor is actually used once every $2\codelen-2$ clock cycles. This motivates us to find a schedule to merge some of the nodes into a single processing element.

\subsection{Pipelined tree architecture}
\label{sec:arch:pipelined}

\begin{figure}[t]
  \centering
  \centerline{\includegraphics[width=\columnwidth]{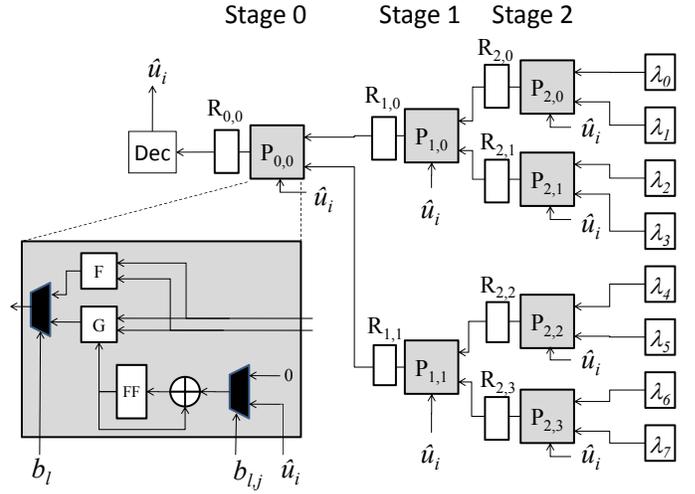}}
  \caption{Pipelined tree SC architecture for $\codelen=8$.}
\label{fig:decoder_pip_tree}
\end{figure}

Further studying of the scheduling reveals
that whenever stage $l$ is activated, only $2^l$ nodes are actually updated. For example, in Figure~\ref{fig:decoder}, when stage 0 is enabled, only one node is updated. Then the $\codelen$ nodes of stage 0 can be implemented using a single processing element (PE). As such, for stage $l$, $2^l$ processing elements are sufficient to update all the nodes. However, this resource sharing does not necessarily guarantee that the memories assigned to the merged nodes can also be merged. Table~\ref{tab:sched_pip_tree} shows the stage activation during the decoding of one vector $\bfy$. When stage $S_l$ is enabled, we indicate which function  ($f$ or $g$) is applied to the $2^l$ activated nodes at stage $S_l$ during each clock cycle (CC). Every generated variable is used twice during the decoding. For example, the four variables generated in stage 2 at CC \#1 are consumed on CC \#2 and CC \#5 in stage 1. This means that, in stage 2, the four registers associated with the $f$ function can be reused at CC \#8 to store the four data values generated by the $g$ function. This observation is applicable to any stage in the decoder. 
The resulting proposed architecture is shown in Figure~\ref{fig:decoder_pip_tree} for $\codelen=8$. The channel LRs, $\lambda_i$, are stored in $\codelen$ registers. The rest of the decoder is composed of a pipelined tree structure that includes $\codelen-1$ PEs, P$_{l,j}$, and $\codelen-1$ registers, R$_{l,j}$ with $0 \leq l \leq \codeloglen-1$ and $0 \leq j \leq 2^l-1$. A decision unit generates the estimated bit $\Hat{u}_i$  which is then broadcast back to every PE. A PE is a configurable element that can perform either the $f$ or the $g$ function. It also includes the $\Hat{u}_s$ computation block that updates the $\Hat{u}_s$ value with the last decoded bit $\Hat{u}_i$ only if the control bit $b_{l,j}=1$. Another control bit is used to select the $f$ or $g$ function. Compared to the butterfly-based structure, the pipelined tree architecture performs the same amount of computation with the same scheduling (see Table~\ref{tab:sched_pip_tree}) but with a smaller number of PEs and registers. The throughput is then the same as in \eqref{eq:throughput} and the decoder has lower hardware complexity
\begin{equation}
  C_{tree}=(\codelen-1) (C_{PE} + \registerComplexity) + \codelen \registerComplexity,
\end{equation}
where $C_{PE}$ represents the complexity of a single PE. In addition to the lower complexity, one can notice that the routing network in the decoder is much simpler in the tree architecture than in the butterfly-based structure. Connections between PEs are also local. This lowers the risk of congestion during the wire routing phase of an integrated circuit design and potentially increases the clock frequency and the throughput.

\begin{table}[t] \centering
  \begin{tabular}{|x{3mm}||@{\hspace{1.4mm}}c@{\hspace{1.4mm}}|@{\hspace{1.4mm}}c@{\hspace{1.4mm}}|@{\hspace{0.9mm}}c@{\hspace{0.9mm}}|@{\hspace{0.9mm}}c@{\hspace{0.9mm}}|@{\hspace{1.4mm}}c@{\hspace{1.4mm}}|@{\hspace{0.9mm}}c@{\hspace{0.9mm}}|@{\hspace{0.9mm}}c@{\hspace{0.9mm}}|@{\hspace{1.4mm}}c@{\hspace{1.4mm}}|@{\hspace{1.4mm}}c@{\hspace{1.4mm}}|@{\hspace{0.9mm}}c@{\hspace{0.9mm}}|@{\hspace{0.9mm}}c@{\hspace{0.9mm}}|@{\hspace{1.4mm}}c@{\hspace{1.4mm}}|@{\hspace{0.9mm}}c@{\hspace{0.9mm}}|@{\hspace{0.9mm}}c@{\hspace{0.9mm}}|}
  \hline
  CC&1&2&3&4&5&6&7&8&9&10&11&12&13&14\tn
  \hline
  \hline
  $S_2$&$f$& & & & & & &$g$& & & & & &\tn \hline
  $S_1$& &$f$& & &$g$& & & &$f$& & &$g$& &\tn \hline
  $S_0$& & &$f$&$g$& &$f$&$g$& & &$f$&$g$& &$f$&$g$\tn \hline \hline
  $\Hat{u}_i$& & &$\Hat{u}_0$&$\Hat{u}_1$& &$\Hat{u}_2$&$\Hat{u}_3$& & &$\Hat{u}_4$&$\Hat{u}_5$& &$\Hat{u}_6$&$\Hat{u}_7$\tn \hline
  \end{tabular}
\caption{Schedule for the butterfly-based and pipeline tree SC architectures ($\codelen=8$).}
\label{tab:sched_pip_tree}
\end{table}

\subsection{Line architecture}
\label{sec:arch:line}
\begin{figure}[t]
  \centering
  \centerline{\includegraphics[width=\columnwidth]{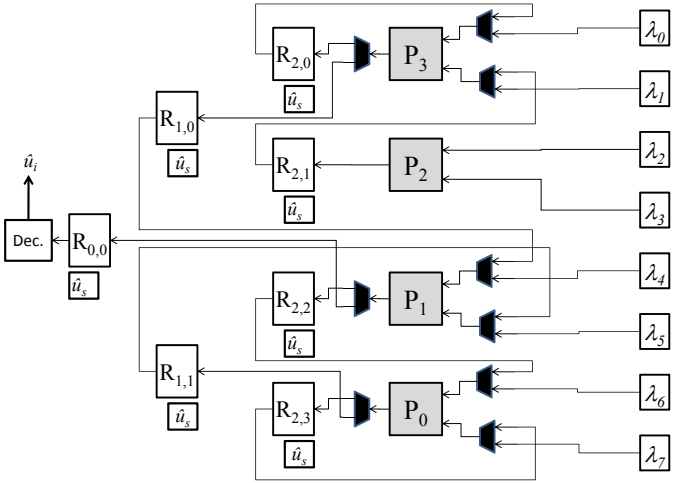}}
  \caption{Line SC architecture for $\codelen=8$.}
\label{fig:line_decoder}
\end{figure}

Despite the low complexity of the pipelined tree architecture, it is possible to further reduce the number of PEs. Looking at Table~\ref{tab:sched_pip_tree}, it appears that only one stage is activated at a time. In the worst case---stage $\codeloglen-1$ is activated---$\frac{\codelen}{2}$ PEs have to be used simultaneously. This means that the same throughput can be achieved with only $\frac{\codelen}{2}$ PEs. The resulting architecture is shown in Figure~\ref{fig:line_decoder} for $\codelen=8$. The processing elements P$_j$ are arranged in a line; while the registers retain a tree structure emulated by a multiplexing resources connecting the two.

For example, since P$_{2,0}$ and P$_{1,0}$ (in Figure~\ref{fig:decoder_pip_tree}) are merged into P$_2$ (in Figure~\ref{fig:line_decoder}), P$_2$ should write either to R$_{2,0}$ or R$_{1,0}$; and it should also read from the channel registers or from R$_{2,0}$ and R$_{2,1}$. The $\Hat{u}_s$ computation block is moved out of P$_j$ and kept close to the associated register because $\Hat{u}_s$ should also be forwarded to the PE. The overall complexity of the line SC architecture is

\begin{equation}
  C_{line}=(\codelen-1) (\registerComplexity + C_{\Hat{u}_s}) + \frac{\codelen}{2} C_{PE} + \left(\frac{\codelen}{2}-1\right)3\muxComplexity+\codelen \registerComplexity,
\label{eq:complexity_line_decoder}
\end{equation}

where $\muxComplexity$ represents the complexity of a 2-input multiplexer and $C_{\Hat{u}_s}$ is the complexity of the $\Hat{u}_s$ computation block. Despite the extra multiplexing logic required to route the data through the PE line, the savings in number of PEs makes this SC decoder less complex than the pipelined tree architecture while achieving the same throughput computed in \eqref{eq:throughput}. The control logic is not included in the complexity estimation since it is negligible compared to processing and memory. This will be confirmed by logic synthesis results in section \ref{sec:hw:asic}.

The Line SC architecture can be seen as a tree architecture in which complexity is reduced by merging some of the PEs without affecting throughput.

\subsection{The min-sum approximation}
\label{sec:min_sum}

SC decoding was originally proposed in the likelihood ratio domain, in which the update rules $f$ and $g$ require multiplications and divisions. Since the cost of implementing these operations in hardware is very high, they are usually avoided in practice. We thus propose to perform SC decoding in the logarithm domain in order to reduce the complexity of the $f$ and $g$ computation blocks. We assume that the channel information is available as log-likelihood ratios (LLRs) $L_i$, which leads to the following alternative representation for equations $f$ and $g$:

\begin{equation}
  \label{eqn:llr_fg}
  \begin{split}
    f(L_a,L_b) &= 2\tanh ^{-1} \left ( \tanh \left (\frac{L_a}{2}\right ) \tanh \left (\frac{L_b}{2}\right ) \right ) \; \text{and}\\
    g_{\Hat{u}_s}(L_a,L_b) &= L_a(-1)^{\Hat{u}_s}+L_b.
  \end{split}
\end{equation}

In terms of hardware implementation, $g$ can easily be mapped to an adder-subtractor controlled by the bit $\Hat{u}_s$. However, $f$ involves some transcendental functions that are complex to implement in hardware. One can notice that the $f$ and $g$ functions are identical to the update rules used in BP decoding of low-density parity-check (LDPC) codes. Consequently, an approximation used in LDPC decoder implementations \cite{MS} can be used to approximate $f$ using the \emph{minimum} function, such that

\begin{equation}
  \label{eqn:llr_fg_approx}
  \begin{split}
    f(L_a,L_b) &\approx \textrm{ sign}(L_a)  \textrm{ sign}(L_b) \min (|L_a|,|L_b|) \; \text{and} \\
    g_{\Hat{u}_s}(L_a,L_b) &= L_a(-1)^{\Hat{u}_s}+L_b.
  \end{split}
\end{equation}

\begin{figure}[t]
  \centering
  \centerline{\includegraphics[width=\columnwidth]{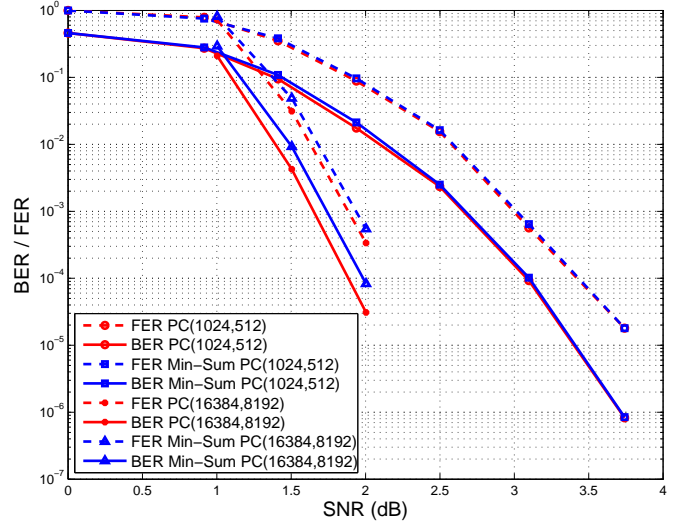}}
  \caption{The min-sum approximation error-correction performance change for PC(1024,512) and PC(16384,8192).}
\label{fig:MS_perf}
\end{figure}

In order to estimate the performance degradation incurred by this approximation, we simulated the performance of different polar codes on an additive white-Gaussian (AWGN) channel with binary phase-shift keying (BPSK). As it can be seen in Figure~\ref{fig:MS_perf}, the performance degradation is minor for moderate code lengths and is very small (0.1dB) for longer codes.

\section{Line decoder hardware implementations}
\label{sec:hw}

Section~\ref{sec:sc-arch} showed that the line architecture has a lower hardware complexity---and is thus more efficient---than its tree-based counterpart. This section presents details and synthesis results of an implementation of the line architecture.

\subsection{Fixed-point simulations}
\label{sec:hw:fp-simu}

\begin{figure}[t]
  \centering
  \centerline{\includegraphics[width=\columnwidth]{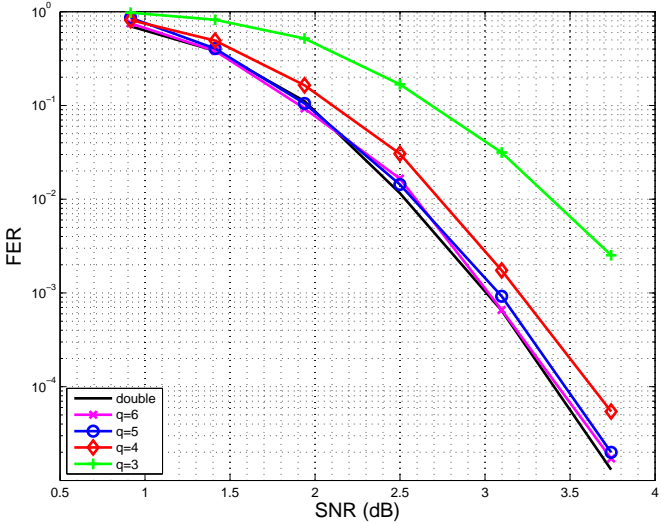}}
  \caption{ Fixed-point FER simulation for PC(1024,512) on AWGN channel. Input saturation $= 3 \sigma$.  }
\label{fig:quantization}
\end{figure}

The number of quantization bits impacts both the decoding performance of the algorithm and the hardware complexity of the decoder. Consequently, prior to implementing the line decoder, a detailed analysis was carried out on a software-based SC decoder in order to find the best tradeoff between performance and complexity. The resulting simulations revealed that fixed-point operations on a limited number of quantization bits attained a decoding performance very similar to that of a floating point algorithm. Figure~\ref{fig:quantization} illustrates this phenomenon for a $\mathtt{PC}(1024,512)$ decoder. It shows that 5 or 6 quantization bits are sufficient to reach near-floating point performance at a saturation level of $\pm 3 \sigma$, which exhibits good performance over all quantization levels. It should be noted that the channel saturation level has a high impact on the performance of low quantization ($q=3, 4$) decoders. The selected saturation value ($\pm 3\sigma$) was chosen from further software simulations not shown here. 

\subsection{Line decoder detailed architecture}
\label{sec:hw:line-archi}

\subsubsection{Processing elements}
\label{sec:hw:line-archi:pe}

The processing element is the main arithmetic component of the line decoder. It embodies the arithmetic logic needed to perform both $f$ and $g$ functions within a single logic component. This grouping, motivated by the fact that all stages of the decoding graph either perform function $f$ or $g$ at any given time, allowed a greater level of resource sharing. PEs also implement the min-sum approximation described in Section~\ref{sec:min_sum}, which allows for much simpler decoding logic, as it replaces three transcendental functions with a single comparator. Since processing elements are replicated $n/2$ times, Equation~\eqref{eq:complexity_line_decoder}, this approximation has a significant impact on the overall size of the decoder.

In this work, processing elements are fully combinational and operate on quantized sign-and-magnitude (SM) coded LLRs. We initially implemented our PEs in two's complement format (TC), for its wide support in HDL languages, but logic synthesis showed a $20\%$ area reduction when using SM instead. Indeed, Equation~\eqref{eqn:llr_fg_approx} shows that the main operations performed on LLRs are addition, subtraction, absolute value, sign retrieval, and minimum value, all of which are very low complexity operations when using SM format.

\begin{figure}[t]
  \centering
  \centerline{\includegraphics[width=\columnwidth]{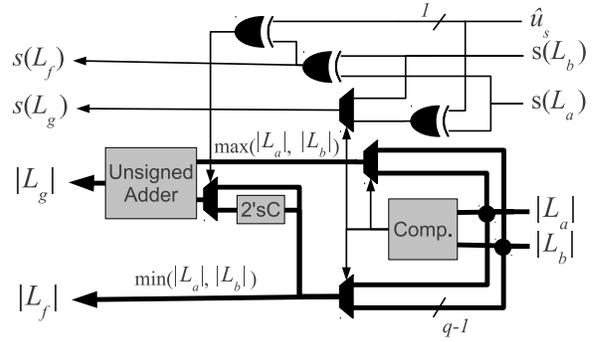}}
  \caption{Processing element architecture.}
\label{fig:pe_architecture}
\end{figure}

Figure~\ref{fig:pe_architecture} illustrates the overall architecture of our SM-based PE. In this figure, $L_a$ and $L_b$ are the two $q$-bit input LLRs of functions $f$ and $g$; a partial sum signal $\hat{u}_s$ controls the behavior of $g$; the sign $s(.)$ and the magnitude $|.|$ of input LLRs are directly extracted; and the comparator is shared for the computation of $|L_f|$, $|L_g|$ and $s(L_g)$. Thick lines and thin lines represent magnitude and sign data paths, respectively.

\subsubsection{Register banks}
\label{sec:hw:line-archi:registers}

As seen in Section~\ref{sec:arch:line}, memory resources are needed to store partial results during the decoding process. The decoder is implemented using two separate memories: one for partial LLR calculations, and another for the partial sums $\hat{u}_s$. The line decoder memory has a tree structure and uses $(2n-1)$ $q$-bit memory cells to store LLRs, in addition to $(n-1)$ $1$-bit cells to store the partial sums $\hat{u}_s$ used to carry out function $g$.

The LLRs memory can be seen as $(\log_2n+1)$ separate memories---one for each stage---with each stage $l$ requiring $2^l$ $q$-bit memory cells. Stage $(\log_2n+1)$ is special in that it contains the received channel LLRs, requiring shift-register capabilities. Each stage produces half as much data as it consumes. This data is written into memory locations read by the subsequent stage, $l-1$.

The partial sum memory combines the $n \log_2n$ partial sums of the decoding graph into $n-1$ memory cells by time-multiplexing each memory cell for use by multiple nodes of the graph.

\subsubsection{Multiplexing}
\label{sec:hw:line-archi:mux}

The shared nature of the processing elements used in the line architecture requires the multiplexing of their inputs and outputs. As shown in Figure~\ref{fig:line_decoder}, memory is implemented using registers, and separate networks of multiplexers and demultiplexers are used to provide them with appropriate inputs from memory and store their outputs to memory, respectively.

An alternate design for the line architecture could make use of SRAM blocks, in which case the multiplexing networks could be avoided completely, as equivalent logic would be directly embodied in the memory decoder of the SRAM modules. This would allow for a more compact memory block, although potentially increasing access time. An even more optimized design could mix both SRAM and registers: looking at table \ref{tab:sched_pip_tree}, it appears that some of the memory elements are accessed more often than others. It would be more efficient to implement these frequently accessed memory elements into registers while keeping the SRAM blocks for less frequently accessed data. In this work, since we target moderate length codes, we choose to use registers  only.

\subsubsection{General control}
\label{sec:hw:line-archi:gen-ctrl}

The line decoder is a multi-stage design which sequentially decodes each codeword. It uses specific control signals \footnote{Binary representations of integers are assumed to be stored in little-endian format} to orchestrate the decoding.

Those control signals are combinational functions of $i$, the current decoded bit number, and $l$, the current stage. These two signals are in turn generated using counters, and some extra logic. The underlying understanding is that up to $\log_2(n)$ stages must be activated in sequence to decode each bit $\hat{u}_i$. Once it has been decoded, this bit is stored in specific partial sums $\hat{u}_s$ for the decoding of subsequent bits, according to the data dependencies highlighted previously.

Both $i$ and $l$ can be viewed as counters, where $i$ counts up from $0$ to $n-1$, for each decoded bit; while $l$ counts down to $0$, from a value between $1$ and $(\log_2(n)-1)$, for each stage. The decoding of a codeword takes $2n-2$ clock cycles overall, as demonstrated in Equation~\eqref{eq:ncc}.

Counter $l$, unlike $i$, is not reset to a fixed value. By making use of the partial computations stored in the LLR memory, it can be reset to the result of a find-first-bit-set ($\mathtt{ffs}$) operation on $i$, corresponding to a modified priority encoder. Specifically, it is reset to $\mathtt{ffs}^\star(i+1)$ upon reaching $0$, according to Equation~\eqref{eqn:fstar}. 

\begin{equation}
\label{eqn:fstar}
\mathtt{ffs}^\star(x_{m-1} \ldots x_1 x_0) =
\left\{
	\begin{array}{ll}
		\min(i) : x_i = 1  & \mbox{   if } x > 0 \\
		m - 1 & \mbox{   if } x = 0
	\end{array}
\right.
\end{equation}

Another control signal deals with the function that the processing elements must perform on behalf of a specific stage. Since the nodes of a given stage all perform the same function at any given time, this signal can be used to control all the PEs of the line. The function selection is performed using Equation~\eqref{eqn:fg_hw}.

\begin{equation}
\label{eqn:fg_hw}
\mathtt{selector}_{f,g}(i_m \ldots i_1 i_0,~l) =
\left\{
	\begin{array}{ll}
		\mathtt{f}  & \mbox{   if } i_l = 0 \\
		\mathtt{g} & \mbox{   if } i_l = 1
	\end{array}
\right.
\end{equation}

\subsubsection{Memory control}
\label{sec:hw:line-archi:mem-ctrl}

Both the LLR and the partial sum memory require significant multiplexing in order to route the proper values from both memories to the PEs, and vice versa.

The multiplexer network mapping the inputs of the processing elements to the LLR memory use the mapping shown in Equation~\eqref{eqn:hw_map_llr_mem_in},

\begin{equation}
\label{eqn:hw_map_llr_mem_in}
\mathtt{MAP}_{\mathtt{LLR}}^{\mathtt{MEM} \rightarrow \mathtt{PE}}(l, p) =
\left\{
	\begin{array}{ll}
		\mathtt{MEM}_{\mathtt{LLR}}(2n - 2^{l+2} + 2p)  & \mbox{   for } L_a \\
		\mathtt{MEM}_{\mathtt{LLR}}(2n - 2^{l+2} + 2p + 1) & \mbox{   for } L_b
	\end{array}
\right.
\end{equation}

where $0 \leq p \leq (\frac{n}{2}-1)$ is the index of the PE in the line. This mapping assumes that the original codeword is stored in memory $\mathtt{MEM}_{\mathtt{LLR}}(0:n-1)$.

The resulting computation is then stored according to the mapping shown in Equation~\eqref{eqn:hw_map_llr_mem_out}, noting that only the first $2^l$ PEs of the line are active in stage $l$.

\begin{equation}
\label{eqn:hw_map_llr_mem_out}
\mathtt{MAP}_{\mathtt{LLR}}^{\mathtt{PE} \rightarrow \mathtt{MEM}}(l, p) =
		\mathtt{MEM}_{\mathtt{LLR}}(2n - 2^{l+1} + p)
\end{equation}

Once stage $0$ has been activated, the output of $\mathtt{PE}_0$ contains the LLR of the decoded bit $i$, and a hard decision $\hat{u}_i$ can be obtained from this soft output using Equation~\eqref{eq:pc-decision}; in other words, if $\mathtt{sign}(\mathtt{LLR}) = 0$. At this point, if bit $i$ is known to be a frozen bit, the output of the decoder is forced to $\hat{u}_i = 0$.
  
Once bit $\hat{u}_i$ has been decoded, this value must be reflected in the partial (modulo-2) sums $\hat{u}_s$ of the decoding graph. Algorithm~\ref{alg:hw_us_update} determines, for each $g$ node with index $z$ in stage $l$, whether it must be updated.

\begin{algorithm}
\caption{Partial sums updating algorithm}
\begin{algorithmic}
\label{alg:hw_us_update}
\STATE $z^\star \gets \mathtt{bitreverse}(z)$
\IF {$l_i = 0$}
    \IF {$l = m-1 \mathtt{~or~} i_{(m-1):(l+1)} = z^\star_{(m-2):l}$}
        \IF {$(l = 0) \mathtt{~or~} ((\mathtt{not}(i_{l:0}) \mathtt{~and~} z^\star_{l:0}) = 0)$}
        	\STATE Node $z$ updates its partial sum with $\hat{u}_i$
        \ENDIF
    \ENDIF
\ENDIF
\end{algorithmic}
\end{algorithm}

One can note that the original decoding graph contains $\frac{n}{2} \log_2(n)$ such partial sums, but that a maximum of $n-1$ of them are used for the decoding of any given bit. With some careful time-multiplexing, it is thus possible to reduce the number of memory cells used to hold the partial sums to $n-1$, a clear reduction in complexity. This is the approach taken in this paper.

Finally, the mapping shown in Equation~\eqref{eqn:mapping_us} connects the partial sum input of $\mathtt{PE}_p$ to the partial sums memory.

\begin{equation}
\label{eqn:mapping_us}
\mathtt{MAP}_{\hat{u}_s}^{\mathtt{MEM}_{\hat{u}_s} \rightarrow \mathtt{PE}_p}(l, p) =
		\mathtt{MEM}_{\hat{u}_s}(n - 2^{l+1} + p)
\end{equation}

All of these mapping equations, together with Algorithm~\ref{alg:hw_us_update}, are efficiently implemented with combinational logic.

\subsection{ASIC synthesis results}
\label{sec:hw:asic}

%\begin{figure}[ht]
%\centering
%\subfigure[Area \textit{vs} Quantization]{
%\includegraphics[width=\columnwidth]{fig/line_quantization.eps}
%\label{fig:area1}
%}
%\subfigure[Area repartition]{
%\includegraphics[width=\columnwidth]{fig/line_decoder_area_repartition.eps}
%\label{fig:area2}
%}
%\label{fig:area}
%\caption[]{Line decoder area, TSMC 65nm, f=500MHz}
%\end{figure}

In order to evaluate the silicon footprint of the line decoder, a generic RTL description of the architecture was designed, and synthesized using a standard cell library. 

This generic description enabled us to generate specific line decoder instances for any code length $n$, code rate $R$, target signal-to-noise ratio SNR, and quantization level $q$. Syntheses were carried out to measure the impact of these parameters on area, using \emph{Cadence RTL Compiler v9.1} and the TSMC 65nm worst case \footnote{Nominal supply voltage and temperature are $V_{dd}=0.9V$ and $T=125^{o}C$, respectively} CMOS standard cell library. Synthesis was driven by Physical Layout Estimators (PLE), which allow a more accurate estimation of interconnection delays and area, compared to the classical wire-load model. The target frequency was set to 500MHz.

\begin{figure}[t]
  \centering
  \centerline{\includegraphics[width=\columnwidth]{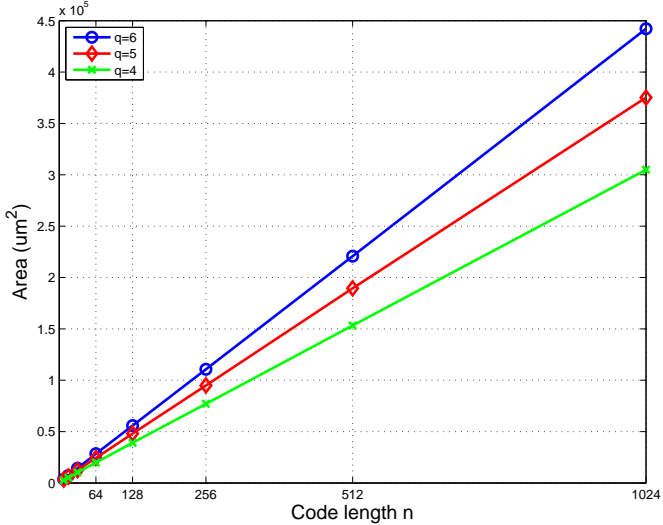}}
  \label{fig:area1}
  \caption{Line decoder area for different quantization and code lengths, TSMC 65nm, f=500MHz}
\end{figure}

A first set of decoders was generated for $8 \leq n \leq 1024$ and $4 \leq q \leq 6$. Figure~\ref{fig:area1} shows the evolution of area as code size and quantization increase. As expected, area grows linearly with $n$ and $q$. The linear hardware complexity of the line decoder validates Equation~\eqref{eq:complexity_line_decoder}.

Then, a second set of decoders was generated and synthesized for $n=1024$, for different codes rates. Synthesis results confirmed that the code rate does not impact hardware complexity. This was expected because the frozen bits are stored in a ROM, whose size is constant; only its contents changes, according to the code rate and target SNR.

\begin{figure}[t]
  \centering
  \centerline{\includegraphics[width=\columnwidth]{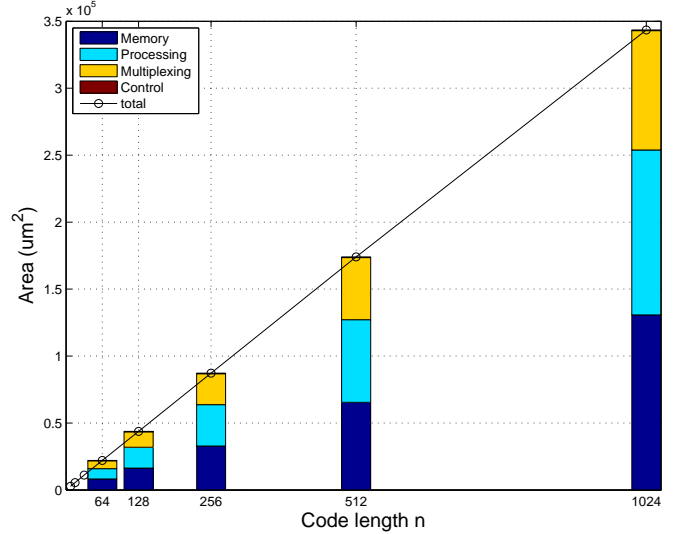}}
  \label{fig:area2}
  \caption{Line decoder area repartition for different code lengths and $q=6$, TSMC 65nm, f=500MHz}
\end{figure}

Finally, a set of decoders was generated for $8 \leq n \leq 1024$ and $q=5$. The area of each component block was extracted, in order to estimate their relative complexity share inside the decoder. Results are shown in Figure~\ref{fig:area2}. Memory resources (register banks), processing logic, and multiplexing, represent $38\%$, $36\%$, and $26\%$ of the total area, respectively. The control logic is negligible ($<1\%$), which is expected as it grows logarithmically in $n$.

\subsection{Verification}
\label{sec:hw-verif}
Verification of the hardware design was carried out by means of functional simulation. Specifically, a testbench was written to exercise the decoder using $10^3$ to $10^6$ randomly-generated noisy input vectors. The output of the simulated hardware decoder was then compared to its software counterpart, whose error-correction capabilities had previously been verified experimentally. This validation was repeated for various combinations of SNR and code lengths to ensure good test coverage.

\section{Conclusion}
\label{sec:conclusion}
Polar codes have recently generated great interest from a theoretical point of view. In this paper, we explore the hardware implementation of polar code decoders; we propose two SC decoders architectures with linear complexity. Software simulations allowed us to validate the proposed min-sum approximation, and to determine implementation parameters, such as the quantization level. For the most efficient decoder---the line-decoder---we provided a detailed description of each component block. Logic synthesis using a standard cell library confirmed the linear evolution of hardware complexity with respect to the code length. 

\bibliographystyle{IEEEbib}      % mathematics and physical sciences
\bibliography{refs}   % name your BibTeX data base

\begin{thebibliography}{1}

\bibitem{channel_polarization}
E.~Arikan,
\newblock ``Channel polarization: A method for constructing capacity-achieving
  codes for symmetric binary-input memoryless channels,''
\newblock {\em IEEE Trans. on Inform. Theory}, vol. 55, no. 7, pp. 3051 --3073,
  Jul. 2009.

\bibitem{TalVardy}
I.~Tal and A.~Vardy,
\newblock ``How to construct polar codes,''
\newblock {\em submitted to IEEE Trans. Inform. Theory, available online as
  \textup{\texttt{arXiv:1105.6164v2}}}, 2011.

\bibitem{Sasoglu2009}
E.~Sasoglu, E.~Telatar, and E.~Arikan,
\newblock ``Polarization for arbitrary discrete memoryless channels,''
\newblock in {\em Proc. IEEE Information Theory Workshop ITW 2009}, 2009, pp.
  144--148.

\bibitem{PC_wiretap}
H.~Mahdavifar and A.~Vardy,
\newblock ``Achieving the secrecy capacity of wiretap channels using polar
  codes,''
\newblock in {\em IEEE ISIT 2010}, Jun. 2010, pp. 913 --917.

\bibitem{perf_PC}
N.~Hussami, R.~Urbanke, and S.B. Korada,
\newblock ``Performance of polar codes for channel and source coding,''
\newblock in {\em IEEE ISIT 2009}, Jun. 2009, pp. 1488 --1492.

\bibitem{tal2011}
Ido Tal and Alexander Vardy,
\newblock ``List decoding of polar codes,''
\newblock in {\em Proc. (ISIT) Symp. IEEE Int Information Theory}, 2011.

\bibitem{Gallager68}
R.~Gallager,
\newblock {\em Information Theory and Reliable Communications},
\newblock John Wiley, New York, 1968.

\bibitem{MS}
M.P.C. Fossorier, M.~Mihaljevic, and H.~Imai,
\newblock ``Reduced complexity iterative decoding of low-density parity check
  codes based on belief propagation,''
\newblock {\em IEEE Trans. on Comm.}, vol. 47, no. 5, pp. 673 --680, May. 1999.

\end{thebibliography}

\end{document}